\documentclass[twocolumn]{svjour3}         
\smartqed  
\usepackage{graphicx}
\begin{document}

\title{Understanding the chemical complexity in Circumstellar Envelopes of C-Rich AGB
stars: The case of IRC +10216}
\subtitle{}

\titlerunning{Chemical complexity in C-rich Circumstellar Envelopes}        

\author{M. Ag\'undez, J. Cernicharo, J. R. Pardo, J. P. Fonfr\'{i}a Exp\'osito \and
        M. Gu\'elin                                                            \and
        E. D. Tenenbaum, L. M. Ziurys, A. J. Apponi
        }

\authorrunning{M. Ag\'undez et al.} 

\institute{M. Ag\'undez, J. Cernicharo, J. R. Pardo, J. P. Fonfr\'{i}a Exp\'osito\at
              Departamento de Astrof\'isica Molecular e Infrarroja, Instituto de Estructura de la Materia, CSIC, Serrano 121, E--28006, Madrid,
              Spain\\
              Tel.: +34 915616800\\
              Fax:  +34 915645557\\
              \email{marce@damir.iem.csic.es}           
           \and
           M. Gu\'elin \at
              Institut de Radioastronomie Millim\'etrique, 300 rue de la Piscine, F--38406 St. Martin d'H\'eres, France
           \and
           E. D. Tenenbaum, L. M. Ziurys, A. J. Apponi \at
              Departments of Chemistry and Astronomy, University of Arizona, 933 North Cherry Avenue, Tucson, AZ 85721
              }
\date{Received: date / Accepted: date}

\maketitle

\begin{abstract}
The circumstellar envelopes of carbon-rich AGB stars show a
chemical complexity that is exemplified by the prototypical object
IRC +10216, in which about 60 different molecules have been
detected to date. Most of these species are carbon chains of the
type C$_n$H, C$_n$H$_2$, C$_n$N, HC$_n$N. We present the detection
of new species (CH$_2$CHCN, CH$_2$CN, H$_2$CS, CH$_3$CCH and
C$_3$O) achieved thanks to the systematic observation of the full
3 mm window with the IRAM 30m telescope plus some ARO 12m
observations. All these species, known to exist in the
interstellar medium, are detected for the first time in a
circumstellar envelope around an AGB star. These five molecules
are most likely formed in the outer expanding envelope rather than
in the stellar photosphere. A pure gas phase chemical model of the
circumstellar envelope is reasonably successful in explaining the
derived abundances, and additionally allows to elucidate the
chemical formation routes and to predict the spatial distribution
of the detected species.

\keywords{astrochemistry \and circumstellar matter \and molecular
processes \and AGB stars \and IRC +10216}
\end{abstract}

\section{Introduction}
\label{sec-intro}

\begin{figure*}
\includegraphics[angle=-90,scale=.68]{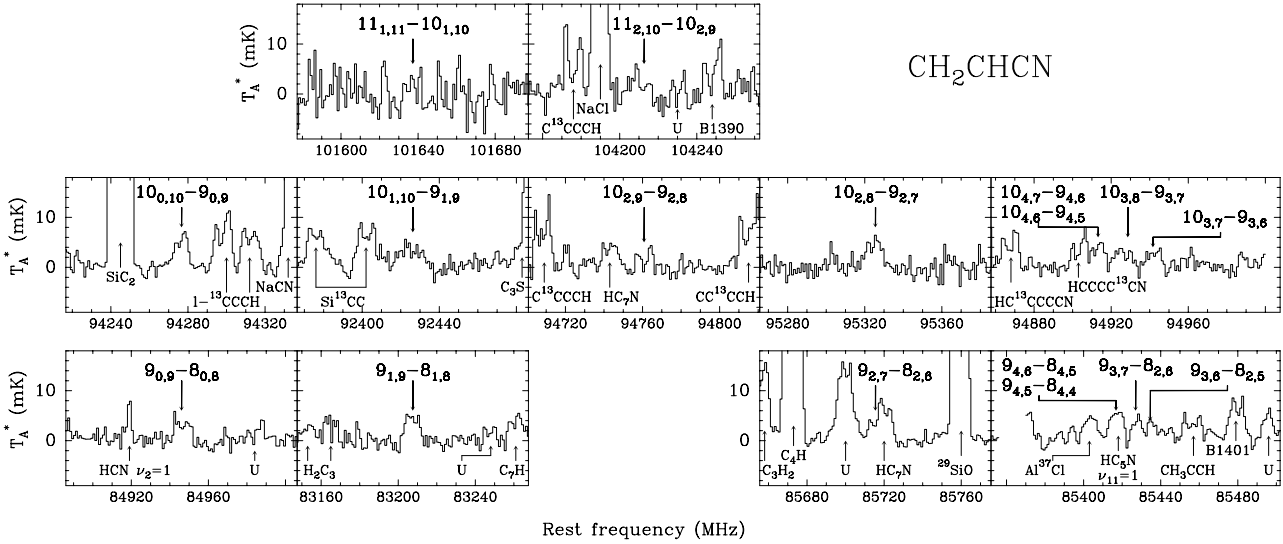}
\caption{Rotational lines of CH$_2$CHCN detected toward IRC +10216
with the IRAM 30m telescope.} \label{fig-ch2chcn-lines}
\end{figure*}

IRC +10216 was discovered in the late sixties as an extremely
bright object in the mid infrared (Becklin et al. 1969). Since
then and with the development of radioastronomy during the
seventies it was recognized as one of the richest molecular
sources in the sky together with some others such as the Orion
nebula, Sagittarius-B2 and the Taurus molecular cloud complex.

IRC +10216 consists of a central carbon-rich AGB star, i.e.
C/O$>$1 in the photosphere, losing mass at a rate of
2-4$\times$10$^{-5}$ M$_{\odot}$ yr$^{-1}$ in the form of a
quasi-spherical wind that produces an extended circumstellar
envelope (CSE) from which most of the molecular emission arises.
To date, some 60 different molecules have been detected in this
source, most of which are organic molecules consisting of a linear
and highly unsaturated backbone of carbon atoms. Among these
species there are cyanopolyynes (HC$_{2n+1}$N) and their radicals
(C$_{2n+1}$N), polyyne radicals (C$_n$H), carbenes (H$_2$C$_n$),
radicals (HC$_{2n}$N) as well as S-bearing (C$_n$S) and Si-bearing
(SiC$_n$) species (Glassgold 1996; Cernicharo et al. 2000).

It is nowadays accepted that the formation of these molecules
occurs either under chemical equilibrium in the dense
(n$>$10$^{10}$ cm$^{-3}$) and hot (T$_k$$\sim$2000 K) vicinity of
the stellar photosphere or in the colder and less dense outer
envelope when the interstellar UV field photodissociate/ionize the
molecules flowing out from the star producing radicals/ions which
undergo rapid neutral-neutral and ion-molecule reactions (Lafont
et al. 1982; Millar et al. 2000). This picture of circumstellar
photochemistry resembles that occurring in cold dense clouds
(T$_k$=10 K, n$\sim$10$^4$ cm$^{-3}$) such as TMC-1 (Kaifu et al.
2004). In both places organic molecules are mostly unsaturated,
which is typical of low temperature non-equilibrium chemistry and
simply reflects the trend of gas phase bimolecular reactions in
ejecting an hydrogen atom and the low reactivity of H$_2$ with
most hydrocarbons.

In this contribution we report on the detection toward IRC +10216
of the partially saturated C-bearing species CH$_2$CHCN, CH$_2$CN
and CH$_3$CCH; the S-bearing molecule H$_2$CS and the
oxygen-carbon chain C$_3$O. All these species are known to exist
in cold dense clouds (Matthews \& Sears 1983; Irvine et al. 1988;
Irvine et al. 1981; Irvine et al. 1989; Brown et al. 1985). Thus,
their detection in IRC +10216 stress the similarity between the
chemistry taking place in cold dense clouds and in the CSEs of
C-rich AGB stars.

\section{Observations}
\label{sec-observ}

The observations of CH$_2$CHCN, CH$_2$CN, H$_2$CS and CH$_3$CCH
were achieved with the IRAM 30m telescope (see e.g.
Fig.~\ref{fig-ch2chcn-lines}) while C$_3$O was observed with both
the IRAM 30m and ARO 12m telescopes.

The IRAM 30m observations were carried out during several sessions
from 1990 to 2005, most of them after 2002 in the context of a
$\lambda$ 3 mm line survey of IRC+10216 from 80 to 115.75 GHz
(Cernicharo et al., in preparation). Two 3 mm SIS receivers with
orthogonal polarizations were used in single-sideband mode with
image sideband rejections $>$20 dB. The standard wobbler switching
mode was used with an offset of 4'. The back end was a 512
two-pole filter with half-power widths and spacings equal to 1.0
MHz.

The ARO 12m observations were done in several runs between 2003
and 2005.  The receivers were dual-channel cooled SIS mixers at 2
and 3 mm, operated in single-sideband mode with $\sim$18 dB
rejection of the image sideband. The back ends were two 256
channel filter banks with 1 and 2 MHz resolutions, configured  in
parallel mode (2$\times$128 channels) for the two receiver
channels. A millimeter auto correlator spectrometer with 2048
channels of 768 kHz resolution was operated simultaneously to
confirm features seen in the filter banks. Data were taken in beam
switching mode with $\pm$2' subreflector throw.

\section{Molecular column densities}
\label{sec-coldens}

\begin{figure}
\includegraphics[angle=-90,scale=.37]{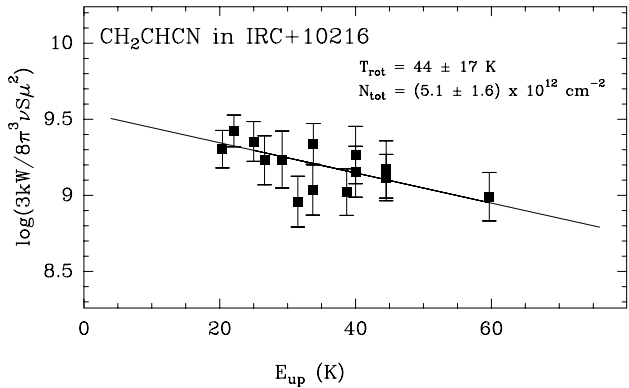}
\caption{Rotational temperature diagram of CH$_2$CHCN.}
\label{fig-ch2chcn-rtd}
\end{figure}

The number of CH$_2$CHCN, CH$_2$CN and C$_3$O lines detected was
large enough to allow us to construct rotational temperature
diagrams (see e.g. Fig.~\ref{fig-ch2chcn-rtd}). The rotational
temperatures derived (see table~\ref{table-coldens}) are within
the range of other shell distributed molecules in IRC +10216:
20-50 K (Cernicharo et al. 2000). For CH$_3$CCH and H$_2$CS we
observed only a few transitions with similar upper level energies
and it was not possible to constrain the rotational temperature
which was assumed to be 30 K.

In Table~\ref{table-coldens} we give the column
densities\footnote{For CH$_2$CN and H$_2$CS, the two molecules
which have two interchangeable nuclei with non zero spin, we have
assumed an ortho-to-para ratio 3:1 when deriving their column
densities.} (averaged over the IRAM 30m beam,
$\theta_{MB}$=21-31'' at $\lambda$ 3 mm) of the species detected
for the first time (in bold) as well as values or upper limits
derived for other related species. The column densities of the new
species are in the range 10$^{12}$-10$^{13}$ cm$^{-2}$. Note for
example that both CH$_3$CN and CH$_3$CCH have very similar column
densities although the lines of CH$_3$CCH are some 30 times weaker
than those of CH$_3$CN mostly because the different electric
dipole moments (0.780 D vs. 3.925 D). The two related species
CH$_3$CN and CH$_2$CN have similar column densities which may
indicate a common chemical origin (see Sec~\ref{sec-chem}). The
carbon-rich nature of IRC +10216 makes oxygen-bearing species to
have a low abundance. For example thioformaldehyde is more
abundant than formaldehyde despite the cosmic abundance of oxygen
being 50 times larger than that of sulphur.

\section{Molecular synthesis in the outer envelope}
\label{sec-chem}

\begin{figure}
\includegraphics[angle=-90,scale=.38]{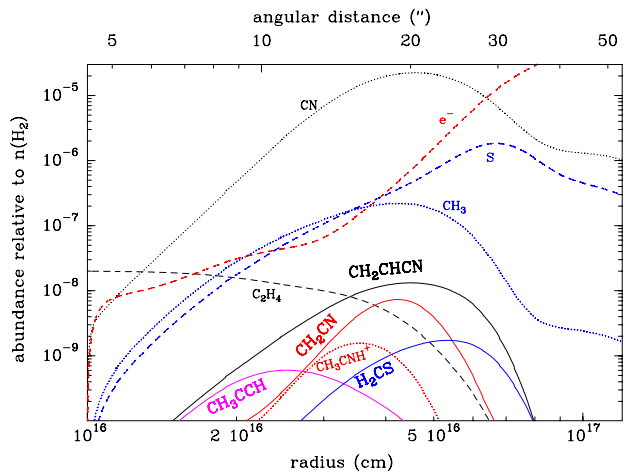}
\caption{Abundances of CH$_2$CHCN, CH$_2$CN, CH$_3$CCH and H$_2$CS
(solid lines) and related species (dotted and dashed lines) given
by the chemical model, as a function of radius (bottom axis) and
angular distance (top axis) for an assumed stellar distance of 150
pc.} \label{fig-cyanides-chem}
\end{figure}

In order to explain how the detected species are formed we have
performed a detailed chemical model of the outer envelope. The
chemical network consists of 385 gas phase species linked by 6547
reactions, whose rate constants have been taken from the
astrochemical databases UMIST99 (Le Teuff et al. 2000) and
osu.2003 (Smith et al. 2004), recently revised\footnote{See
http://www.udfa.net and
http://www.physics.ohio-state.edu/$\sim$eric/research.html}. The
temperature and density radial profiles as well as other physical
parameters are taken from Ag\'undez \& Cernicharo (2006). The
resulting abundance radial profiles for CH$_2$CHCN, CH$_2$CN,
CH$_3$CCH, H$_2$CS and related species are plotted in
Fig.~\ref{fig-cyanides-chem}. The model predicts that these four
molecules together with C$_3$O (not plotted in
Fig.~\ref{fig-cyanides-chem}, see Tenenbaum et al. (2006) for a
detailed discussion) are formed with an extended shell-type
distribution ($r$$\sim$20'') via several gas phase reactions.

Vinyl cyanide (CH$_2$CHCN) is formed by the reaction between CN
and ethylene (C$_2$H$_4$), which is most likely the main formation
route in dark clouds (Herbst \& Leung 1990). Fortunately, the
reaction has been studied in the laboratory and has been found to
be very rapid at low temperature (Sims et al. 1993) and to produce
vinyl cyanide (Choi et al. 2004). The predicted column density
agrees reasonably well with the observational value.

\begin{table}
\caption{Column densities of some molecules in IRC +10216}
\label{table-coldens}
\begin{tabular}{llr@{  }rcc}
\hline\noalign{\smallskip}
\multicolumn{1}{c}{Molecule} & \multicolumn{1}{l}{T$_{rot}$} & \multicolumn{4}{c}{N$_{tot}$}    \\
\multicolumn{1}{c}{} & \multicolumn{1}{l}{(K)} & \multicolumn{4}{c}{(cm$^{-2}$)}    \\
\hline
                    &                     & \multicolumn{2}{c}{observed} & \multicolumn{2}{c}{calculated} \\
\cline{3-6}
                    &                     & & & \multicolumn{1}{c}{This Work} & \multicolumn{1}{c}{MI00} \\
\cline{5-6} \noalign{\smallskip}\hline\noalign{\smallskip}
\textbf{CH$_2$CHCN} & 44                  & 5.1(12)    &      & 1.4(13)       & 2.2(11)     \\
\multicolumn{6}{c}{} \\
CH$_3$CN            & 40                  & 3.0(13)  &        & 6.3(12)       & 6.8(12)     \\
\textbf{CH$_2$CN}   & 49                  & 8.4(12)   &       & 4.6(12)       & 1.4(13)     \\
CH$_3$C$_3$N        & 30$^a$ & $<$1.3(12) &      & 1.2(11)        & 1.4(12)     \\
\textbf{CH$_3$CCH}  & 30$^a$ & 1.6(13)    &      & 1.1(12)       & 8.0(12)     \\
CH$_3$C$_4$H        & 30$^a$ & $<$9.7(12) &      & 8.2(12)       & 9.0(12)     \\
\multicolumn{6}{c}{} \\
H$_2$CO             & 28                  & 5.4(12) & \cite{for04}    & 2.8(12)       & --          \\
\textbf{H$_2$CS}    & 30$^a$ & 1.0(13) &         & 1.3(12)       & 4.4(11)     \\
\multicolumn{6}{c}{} \\
C$_2$O              & 30$^a$ & $<$7.0(12) & \cite{ten06} & 7.0(11)       & --          \\
\textbf{C$_3$O}     & 27                  & 2.6(12) & \cite{ten06}    & 2.8(12)       & --          \\
C$_4$O              & 30$^a$ & $<$6.0(12) & \cite{ten06} & --            & --          \\
C$_5$O              & 30$^a$ & $<$3.0(12) & \cite{ten06} & --            & --          \\
\noalign{\smallskip}\hline
\end{tabular}
Notes: a(b) refers to a$\times$10$^b$. The total column density
across the source N$_{tot}$ is twice the radial column density
N$_{rad}$. A superscript "a" indicates an assumed value.\\
References: (MI00) Chemical model of Millar et al. (2000).
\end{table}

In our model both CH$_2$CN and CH$_3$CN are mostly formed ($>$90
\%) through the dissociative recombination (DR) of CH$_3$CNH$^+$
\begin{displaymath}
\begin{array}{rl}
CH^+ \displaystyle{\stackrel{H_2}{\longrightarrow}} CH_2^+ \displaystyle{\stackrel{H_2}{\longrightarrow}} CH_3^+ \displaystyle{\stackrel{HCN}{\longrightarrow}} CH_3CNH^+ & \displaystyle{\stackrel{e^-}{\longrightarrow}} CH_2CN \\
 & \displaystyle{\stackrel{e^-}{\longrightarrow}} CH_3CN \\
\end{array}
\end{displaymath}
whereas the major destruction process ($>$90 \%) for both species
is photodissociation. The branching ratios in the DR of
CH$_3$CNH$^+$ are not known and are assumed to be equal. Assuming,
as we do, that the photodissociation rate of CH$_3$CN and CH$_2$CN
are equal, the observed CH$_3$CN/CH$_2$CN ratio suggests branching
ratios in the DR of CH$_3$CNH$^+$ of 0.8 for the (CH$_3$CN + H)
channel and 0.2 for (CH$_2$CN + H$_2$ or 2H). This estimate will
be strongly affected if the photodissociation rates of CH$_3$CN
and CH$_2$CN are very different but not if, as has been suggested
by Herbst \& Leung (1990) and Turner et al. (1990), CH$_2$CN does
indeed react with atomic oxygen, the abundance of which is too low
at the radius where CH$_2$CN is present.

The synthesis of CH$_3$CCH involves ion-molecule reactions with
the dissociative recombination of the C$_3$H$_5$$^+$ and
C$_4$H$_5^+$ ions as the last step. The model underproduces it by
an order of magnitude, probably due to uncertainties and/or
incompleteness in the chemical network, which affect the formation
rate of the last step species C$_3$H$_5$$^+$ and C$_4$H$_5^+$. The
heavier chain CH$_3$C$_4$H is also predicted with a column density
even higher than that of CH$_3$CCH. The larger rotational
partition function works against lines being detectable, but the
higher dipole moment (1.21 D vs. 0.78 D) could result in line
intensities similar to those of CH$_3$CCH.

Thioformaldehyde is formed by the reaction S + CH$_3$ and in less
extent through the DR of H$_3$CS$^+$ (see Ag\'undez \& Cernicharo
2006 for a detailed discussion). The order of magnitude of
discrepancy between model and observations reduces to a factor 4
with further non-local non-LTE radiative transfer calculations. We
note, however, that a significant fraction of both H$_2$CO and
H$_2$CS could be formed in grain surfaces by hydrogenation of CO
and CS respectively.

\begin{figure}
\includegraphics[angle=0,scale=.42]{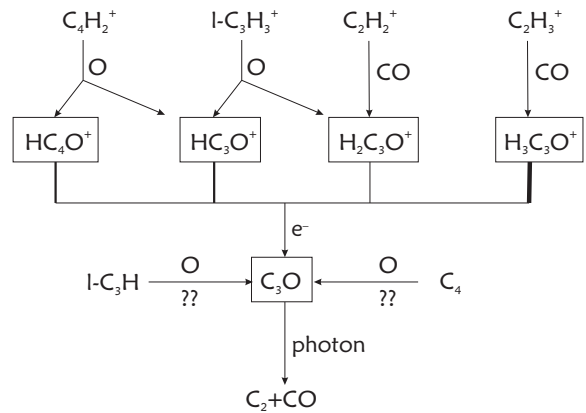}
\caption{Scheme showing the main chemical formation routes to
C$_3$O. From Tenenbaum et al. (2006).} \label{fig-c3o-chem-scheme}
\end{figure}

The detection of C$_3$O in IRC +10216 (Tenenbaum et al. 2006),
only previously detected in the dark clouds TMC-1 (Brown et al.
1985) and Elias 18 (Trigilio et al. 2007), stress both the
similarity between dark clouds and C-rich CSEs chemistries and
also the non-negligible oxygen chemistry taking place in C-rich
CSEs. Actually astrochemical networks consider that C$_3$O is
formed through DR of the molecular ions H$_n$C$_m$O$^+$ (see
Fig.~\ref{fig-c3o-chem-scheme}). However, the C$_3$O column
density derived in IRC +10216 is an order of magnitude higher than
calculated which could imply additional chemical routes for its
formation, e.g. neutral-neutral reactions of atomic oxygen with
carbon chain radicals such as those suggested in
Fig.~\ref{fig-c3o-chem-scheme}.

\begin{acknowledgements}
We would like to thank the IRAM 30m telescope staff for their
assistance during the observations. MA acknowledges a grant from
Spanish MEC: AP2003-4619.
\end{acknowledgements}

\end{document}